\documentclass[preprint,journal]{vgtc}            % preprint (journal style)

\preprinttext{}
\vgtccategory{Research}

\vgtcpapertype{theory/model}

\title{Visualizing Spatial Point Clouds: A Task-Oriented Taxonomy}

\author{%
  Mahsa Partovi, and
  Federico Iuricich.
}

\authorfooter{
  \item
  	Mahsa Partovi is with the School of Computing, Clemson University.
  	E-mail: mpartov@clemson.edu
  \item
  	Federico Iuricich is with the School of Computing, Clemson University.
  	E-mail: fiurici@clemson.edu.

}

\abstract{%
  \lipsum[1] % filler text. Replace with your abstract.
  A free copy of this paper and all supplemental materials are available at \url{https://OSF.IO/2NBSG}.
}


\renewcommand{\manuscriptnotetxt}{}

\graphicspath{{figs/}{figures/}{pictures/}{images/}{./}} % where to search for the images

\usepackage{tabularx}                      % only used for the table example
\usepackage{booktabs}                  % only used for the table example
\usepackage{lipsum}                    % used to generate placeholder text
\usepackage{mwe}                       % used to generate placeholder figures

\usepackage{mathptmx}                  % use matching math font
\usepackage{todonotes}

\usepackage{multirow}

\abstract{The visualization of 3D point cloud data is essential in fields such as autonomous navigation, environmental monitoring, and disaster response, where tasks like object recognition, structural analysis, and spatiotemporal exploration rely on clear and effective visual representation. Despite advancements in AI-driven processing, visualization remains a critical tool for interpreting complex spatial datasets. However, designing effective point cloud visualizations presents significant challenges due to the sparsity, density variations, and scale of the data. In this work, we analyze the design space of spatial point cloud visualization, highlighting a gap in systematically mapping visualization techniques to analytical objectives. We introduce a taxonomy that categorizes four decades of visualization design choices, linking them to fundamental challenges in modern applications. By structuring visualization strategies based on data types, user objectives, and visualization techniques, our framework provides a foundation for advancing more effective, interpretable, and user-centered visualization techniques.%
} % end of abstract

\usepackage{xcolor}
\usepackage[most]{tcolorbox}

\tcbset{
  boxrule=0.4pt,
  boxsep=1pt,
  left=1pt,
  right=1pt,
  top=0.5pt,
  bottom=0.5pt,
  sharp corners,
  enhanced,
  colback=white,
}

\newtcbox{\taskparts}{colframe=blue, colback=blue!10}
\newtcbox{\taskspatial}{colframe=purple, colback=purple!5!white}
\newtcbox{\tasktemporal}{colframe=green!50!black, colback=green!5!white}

\begin{document}


%% The only exception to this rule is the \firstsection command
\firstsection{Introduction}

\maketitle

Spatial point cloud data, generated by technologies such as LiDAR, photogrammetry, and depth cameras, has become indispensable across a range of applications \cite{yue_lidar_2018}\cite{wang_lidar_2018}\cite{bock_visualizationbased_2017}. These datasets provide highly detailed 3D representations of objects and environments, enabling advanced spatial analysis. Despite their potential, the unstructured nature, large scale, and high dimensionality of point cloud data pose significant challenges for effective analysis and interpretation, particularly in tasks that rely on human expertise and decision-making. While visualization research has engaged with point cloud data for over four decades, analyzing these datasets remains a tedious and highly complex task.

The objective of this taxonomy is to systematically map the design space for spatial point cloud visualization, with a strong emphasis on task efficiency. To achieve this, we have reviewed over 100 papers on point cloud visualization, comparing existing visualization techniques with data processing approaches proposed by the machine learning community for point cloud analysis \cite{Guo2021}.

In our review, we critically examine these techniques in relation to visualization tasks, assessing their effectiveness in supporting analytical objectives. To organize this body of knowledge, we adopt the well-established What–Why–How framework \cite{munzner_visualization_2014}, providing a user-centric perspective on point cloud visualization methodologies.

By distinguishing well-established techniques from underexplored design choices, this taxonomy wants to establish a roadmap for future research, encouraging the development of task-driven techniques, evaluations, and comparative studies that establish clear performance benchmarks.

% This survey highlights a key issue in point cloud visualization research: a predominant focus on achieving realism and scalability, often at the expense of task efficiency and effectiveness. As a result, there is an abundance of 3D scatterplot designs aimed at general-purpose rendering of point cloud data that do not really support specific tasks such as object identification or tracking. On the other hand, many other designs have gone unexplored or have not been evaluated for task efficiency \cite.

% The objective of this taxonomy is to systematically delineate the design space for spatial point cloud visualization, emphasizing usability and task efficiency. By distinguishing between well-studied techniques and underexplored design choices, we aim to provide a roadmap for future research. This roadmap encourages the development of task-oriented, comparative evaluations to establish robust rankings and metrics that guide the creation of more effective visualization techniques for point cloud data.

The remainder of this paper is structured as follows. In \Cref{sec:scope}, we define the scope and methodology of our taxonomy study. Following the What–Why–How framework \cite{munzner_visualization_2014}, we first explore data abstraction (What?) in \Cref{sec:data}, focusing on the key characteristics of point cloud data. In \Cref{sec:task}, we analyze task abstraction (Why?), examining the motivations and objectives behind point cloud visualization. \Cref{sec:design} introduces the design choices (How?) for point cloud visualization, including an in-depth discussion of the fundamental visualization primitives: points (\Cref{sec:points}), grids (\Cref{sec:grid}), and geometries (\Cref{sec:geometry}). Finally, in \Cref{sec:discussion}, we summarize key insights from this review and outline potential directions for future research.

\section{Scope and Method}
\label{sec:scope}

The focus of this survey is on spatial point cloud visualization, with a specific emphasis on low-level visual tasks. We prioritize results that have concentrated on evaluating perceptual and cognitive factors in point cloud visualization. However, as will become evident from reading this manuscript, research on this topic remains limited, underscoring the need for further exploration.

Our work complements existing surveys that focus on computational methods for point cloud processing, such as deep learning-based analysis \cite{Guo2021}, or algorithms focusing on registration \cite{huang_comprehensive_2021}, denoising \cite{zhou_point_2022}, completion \cite{Tesema2024}, and segmentation \cite{Nguyen2013}. It also builds upon surveys that review specific visualization techniques, including surface mesh reconstruction \cite{Berger2017}, non-photorealistic rendering \cite{Wegen2024}, and real-time rendering methods \cite{Kivi2022}.

We focus exclusively on spatial point cloud visualization and do not cover techniques related to non-spatial point-based visualization, which have been addressed in other surveys \cite{Sarikaya2018}.

{\bf Methodology.} Our review process began with a systematic selection of articles from major journals specializing in computer graphics and visualization. The primary sources for our literature review included {\em IEEE Transactions on Visualization and Computer Graphics (TVCG)}, {\em Computer Graphics Forum}, {\em IEEE Computer Graphics and Applications}, the {\em Journal of Visualization}, and {\em Transaction on Graphics (TOG)}. We conducted automated searches using the keywords "point cloud rendering," "point cloud visualization," and "point cloud."

To ensure comprehensive coverage, we supplemented these searches with targeted queries on Google Scholar, identifying relevant works on, perceptual studies, and existing taxonomies. Our initial search yielded over 1,200 papers. To refine the selection, we focused on works published between 1990 and 2025, applying additional filtering criteria to exclude publications that were not directly relevant to our study, such as purely theoretical works or algorithm-centric papers lacking a connection to visualization. We complemented this with more than 50 papers focusing on applications involving point clouds.

\section{Data abstraction (What)}
\label{sec:data}

Different definitions and categorizations of point clouds have been proposed based on the application of interest or the acquisition methodology \cite{Wegen2024}. In this survey, we abstract our classification from specific applications.

At its core, a spatial point cloud is a collection of data points defined by at least three spatial coordinates $(x, y, z)$. These coordinates describe the domain of the point cloud. Building on this foundation, we identify four key characteristics defining the data type for spatial points clouds:

\begin{itemize}
    \item {\em Point-Object Relationship} - Describes whether a point represents an independent entity or is a sample of a larger object.  
    \item {\em Number of Objects Sampled} - Specifies whether the dataset represents a single object, or a large-scale environment.
    \item {\em Number of Views} - which describes whether the point cloud is generated from a single viewpoint or aggregated from multiple perspectives.
    \item {\em Dynamicity} - which describes whether the dataset captures static objects or if dynamic elements are involved.
\end{itemize}

{\bf Point-Object Relationship.} The first distinction is about the role of an individual point within a spatial point cloud. In most applications, a point represents a sample of a larger entity rather than a standalone data item. Understanding this distinction is crucial in the visualization design phase, as individual points are rarely informative in isolation. Therefore, effective visualization are expected to emphasize and facilitate the perception of this part-of-a-whole relationship.
However, exceptions exist, particularly in particle simulations and cosmological data analysis, where each point may represent an entire physical entity, such as a planet or a galaxy \cite{Cuesta-Lazaro2024}. In such cases, while still spatially embedded, these points function more like independent data items, similar to points in a scatterplot representation of tabular data \cite{Sarikaya2018}.

{\bf Number of Objects Sampled.} The second characteristic further expands on the part-of-a-whole nature of point clouds by distinguishing between datasets that represent a single object and those that capture an entire scene composed of multiple objects.

A {\em single-object} point cloud focuses on a distinct entity, such as a manufactured part. In contrast, a {\em scene} consists of multiple objects, such as an urban environment or a natural landscape. While both data types can suffer from occlusion issues, the nature and severity of these occlusions, as well as the primary visualization tasks associated with each, differ significantly.

Single-object point clouds are typically used in shape analysis \cite{Berger2017}, or quality inspection \cite{shan_gpa-netno-reference_2024}, where precise geometric detail is paramount. Scene-level point clouds, on the other hand, are more common in applications like autonomous driving \cite{Arief_2019_CVPR_Workshops}, robotics \cite{Nguyen2013}, and architecture \cite{tan_qu_usage_2015}.

{\bf Number of Views.} While this survey focuses on the visualization aspects of point clouds rather than the technical specifics of acquisition devices, the number of viewpoints used during data collection plays a crucial role in defining the data type and, consequently, in determining appropriate visualization strategies.

Point clouds can be classified based on whether they are generated from a single viewpoint or multiple viewpoints. A single-view acquisition produces what can be considered a pseudo-3D point cloud, as it captures only the surfaces visible from the sensor’s perspective. In contrast, a multi-view acquisition integrates data from multiple perspectives, resulting in a more complete 3D representation.

For example, a single LiDAR scan captures only the surfaces directly facing the sensor. This produces a pseudo-3D structure that is closer to a depth map and can often be arranged on a 2D lattice, making 2D visualization approaches a viable option. Conversely, a multi-view scan registers a fully 3D point cloud with scattered points distributed in 3D space \cite{blanc_genuage_2020}.

These differences significantly impact visualization design. In single-view point clouds, occlusions occur during data acquisition, inherently limiting certain tasks such as object recognition or distance estimation (see Section \ref{sec:task}). In contrast, occlusions in multi-viewpoint clouds arise at the visualization stage, where rendering and interaction techniques must address visibility issues to support effective analysis.

{\bf Dynamicity.} The final distinction in point cloud data types concerns whether the content of the point cloud changes. Here, we deliberately avoid using the term "time" in a general sense, as time can also be involved in multi-viewpoint acquisitions, where a moving sensor constructs a full 3D representation over time. Instead, the static vs. dynamic distinction refers specifically to whether the elements in the point cloud remain fixed or exhibit motion \cite{Guo2021}.

A static point cloud assumes that all points remain unchanged. In contrast, a dynamic point cloud captures motion and can involve different types of movement. One type of motion, known as {\em intra-scan motion}, occurs when objects move while the point cloud is being captured. Since most point cloud acquisition methods require a finite time interval to complete a scan, moving elements may appear distorted, much like motion blur in video frames \cite{Pomerleau_Liu_Colas_Siegwart_2012}. Although the resulting dataset may still be considered static (as all points are fixed after capture), the visualization may reveal motion cues, making tasks such as tracking or orientation estimation possible.

A different case arises with {\em inter-scan motion}, where movement occurs between separate acquisitions, resulting in a time-varying point cloud in which each data point is associated with a specific time step. These datasets support dynamic analysis tasks such as object tracking \cite{vaquero_deconvolutional_2017}, and change detection \cite{Tateosian_Mitasova_Thakur_Hardin_Russ_Blundell_2014}.

\subsection{Point cloud attributes}

Spatial point clouds often include additional attributes that provide richer information about the data. These attributes may be directly captured by sensors during acquisition or derived through post-processing. In this survey, we do not distinguish between these sources but instead classify point cloud attributes into two broad categories: {\em categorical} and {\em quantitative}.

{\bf Categorical attributes.} Categorical attributes primarily encode discrete information, such as class labels, which describe either the semantic meaning of individual points or their grouping into clusters.

{\em Class labels} explicitly define the part-to-whole relationships discussed earlier, indicating how multiple points belong to the same object or one of its functional components. For instance, in a point cloud of a vehicle, clustering attributes may differentiate the chassis, windows, and wheels.

{\em Semantic labels}, on the other hand, provide higher-level descriptions of the objects represented by the points (e.g., grass, building, tree). While semantic attributes are useful for general classification, they do not inherently support clustering at finer levels. For example, in a forest point cloud, labeling canopy points as "vegetation" does not directly facilitate the segmentation of individual trees \cite{burt_extracting_2019}.

{\bf Quantitative attributes.} Quantitative attributes can take various forms, including scalar, vector, and multivariate values. 

{\em Scalar attributes} are single-valued numerical properties. Common examples include reflection intensity, which is often recorded by LiDAR sensors, as well as physical measurements such as temperature, pressure, or density in scientific datasets. Another important scalar property in point cloud analysis is curvature, which describes the local geometric characteristics of a surface.

{\em Vector attributes} provide directional information. The most prevalent example is the normal vector, which defines the local orientation of the sampled surface and is critical for surface reconstruction and shading techniques. Other vector attributes include motion vectors, which describe the dynamic properties of moving points.

{\em Multivariate attributes} involve multiple values per point, with color being the most common example. Color attributes can either be acquired directly from RGB cameras or generated through photogrammetry.

\section{Task Abstraction (Why)}
\label{sec:task}

In this section, we generalize domain specific tasks we encountered in our reviewing activity into more general visual tasks that users and domain experts typically perform when interacting with point cloud data. The goal of such abstraction, is to simplify the identification of commonalities between visualization techniques, enabling them to serve similar objectives across different domains \cite{munzner_visualization_2014}.

Although the data visualization literature already presents a broad range of model tasks \cite{Rind_Aigner_Wagner_Miksch_Lammarsch_2016}, we argue that spatial point clouds introduce unique challenges, warranting a dedicated task taxonomy. To develop this, we reviewed existing research and reconciled taxonomies proposed for abstract point data representations \cite{Sarikaya2018} with those for deep learning-based point cloud processing \cite{Guo2021}.

\begin{figure*}[h]
    \centering
    \includegraphics[width=0.95\linewidth]{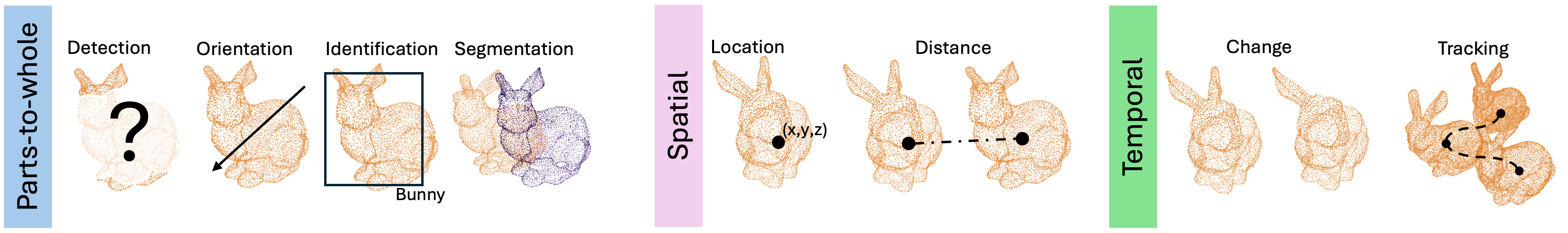} % Replace with your image filename
    \caption{Our taxonomy categorizes low-level user tasks into three main groups: Parts-to-Whole, where the focus is on understanding structural characteristics from a collection of points; Spatial relations, which involves evaluating the spatial relationships between points or objects; and Temporal relations, where the goal is to analyze changes in objects over time.}
    \label{fig:tasks}
\end{figure*}

Our proposed taxonomy organizes user tasks into three primary categories based on the relationships involved: {\em parts-to-whole relations}, {\em spatial relations}, and {\em temporal relations} (see Figure \ref{fig:tasks}). These tasks are all intended as low-level, abstract objectives as defined by Rind et al. \cite{Rind_Aigner_Wagner_Miksch_Lammarsch_2016}.

\subsection{Parts-to-whole relations}
In the analysis of spatial point clouds, a fundamental objective is to effectively identify and group points that belong to the same object or part of a larger scene. In computer vision, this task is often differentiated based on the models used and the desired output, such as centroid locations, bounding boxes, or object segmentation \cite{ blanc_genuage_2020}.

For visual tasks, Johnson’s criterion for analyzing the observer's ability to perform visual tasks provides a helpful framework. It distinguishes four categories: detection (the ability to determine whether an object is present, e.g., "does the point cloud represent X objects?"), orientation (the ability to determine the direction of the object), recognition (the ability to identify the type of object, e.g., "the object is a tree"), and identification (the ability to discern a specific object, e.g., "the tree is a conifer") \cite{1985SPIE..513..761J}.

Our proposed set of low-level tasks integrates these visual categories with small but essential modifications tailored specifically to the characteristics of point cloud data. Namely we propose four tasks that builds upon each other in terms of complexity.

{\bf Detection.} Object detection refers to the ability to perceive the presence or absence of an object in the rendered point cloud and to determine the number of such objects. It is one of the most studied tasks in user-centered point cloud visualization, with research showing its strong link to the user’s ability to visually perceive structure and depth \cite{Aygar_Ware_Rogers_2018, gruchalla_structure_2021}.  This task becomes more challenging in the presence of occlusions or noise, as objects may be hidden or partially obscured \cite{Aygar_Ware_Rogers_2018}. Notably, object detection can be seen as the dual of noise recognition, which plays an equally important role in point cloud analysis. While detection focuses on identifying meaningful structures, noise recognition aims to identify artifacts that may arise from sensor errors, misalignment, or environmental interference. Effective noise identification is essential for data cleaning, as eliminating irrelevant or misleading points can greatly improve the accuracy of subsequent tasks. 

{\bf Orientation.} Object orientation refers to the task of determining the direction or positioning of an object within a scene. It involves understanding how an object is oriented relative to the viewpoint or other objects within the environment. This ability is crucial in applications such as robotics, where accurately perceiving the rotation and alignment of objects is key for navigation and manipulation. While object detection has received more attention in the literature, orientation remains equally important, especially in dynamic point clouds. In these cases, objects and points are often assumed to be moving over time, with their trajectories frequently dictated by changes in object orientation (e.g., the steering of a car while navigating) \cite{Hou2022}.

{\bf Identification.} Object identification refers to the ability to recognize a specific object within a point cloud, often requiring the user to interpret both the object’s visual features and its contextual information within the scene. While Johnson distinguishes between identification and recognition as separate tasks, we find it more fitting to combine them under a single category. Both tasks aim to uncover the semantic meaning of the point cloud data, which is influenced by the point cloud's resolution and the richness of the visual details. This process is closely related to classification tasks in computer vision, where objects are assigned to specific categories. Effective object identification depends heavily on the user's ability to perceive the 3D structure and discern subtle object features \cite{gruchalla_structure_2021}.

{\bf Segmentation.} Object segmentation refers to the task of identifying and delineating physical boundaries between different objects within a point cloud. The goal is to map each point in the point cloud to a specific object or part of an object. Depending on the context, segmentation can refer to different levels of detail: mesh segmentation (dividing a mesh into its functional components) or scene segmentation (separating a scene into discrete objects) \cite{blanc_genuage_2020}. Segmentation is generally considered a difficult task due to factors such as occlusions, redundant points, uneven sampling density, and the lack of explicit structural information in point clouds \cite{Nguyen2013}.

\subsection{Spatial relationships}

The next group of tasks focuses on understanding spatial relationships between points, objects, and their surrounding environment. Specifically, this involves two key tasks: determining object location and assessing relative distances. These tasks are critical to consider when selecting visualization techniques since many visualization methods, such as projections or dimensionality reductions, distort spatial relationships.

{\bf Location.} This task involves visually estimating the three-dimensional position of objects within the original point cloud. The ability to accurately locate objects is strongly tied to spatial context and is of great importance for situational awareness \cite{Tateosian_Mitasova_Thakur_Hardin_Russ_Blundell_2014}. However, determining precise locations in a 3D scene presents challenges when using a 2D display. This limitation can make it difficult for users to orient themselves within the scene and judge the distance between themselves and an object, which is a crucial factor in pinpointing exact locations \cite{Aygar_Ware_Rogers_2018}. The challenge becomes even greater when visualization techniques distort or alter the spatial domain, potentially affecting users' ability to correctly interpret object positions.

% Input is a point cloud representing a scene and any object in the point cloud, and the output is the 3 dimensional position of the object in the scene. While object detection are concerned with semantic knowledge of what are the objects in the point cloud, understand object location focuses on pinpointing where the objects are. This could arise naturally in applications. For example, in autonomous driving or robotics, a user looking at a point cloud visualization might need to know where the cars or people are and how far the next turn is. Not limited to it, in visualization of AGVs moving in a warehouse, the stakeholders would want to know where they are and if there are any collisions. Visual tasks such as these require the users to understand where the objects are in the 3 dimensional scene. 
    
{\bf Distance.} Closely related to object location is the task of assessing distances, which includes both assessing distances between objects and estimating their distance from the observer. Understanding an object's distance from the observer relies heavily on depth perception, requiring users to infer three-dimensional spatial relationships from a two-dimensional visual input \cite{Aygar_Ware_Rogers_2018}. Meanwhile, understanding relative distances between objects involves both qualitative spatial reasoning (e.g., determining which object is closer) and quantitative measurements (e.g., estimating exact distances).

While point clouds provide precise distance measurements that can be leveraged computationally, human interpretation remains crucial, especially in high-stakes scenarios such as rescue missions, where situational awareness is critical \cite{Kot_Novak_Babjak_2016}. Although distance estimation is an independent task, research has shown that it plays a fundamental role in supporting other tasks. For instance, accurately perceiving distances improves structural understanding and enhances the semantic interpretation of point clouds \cite{gruchalla_structure_2021}. Additionally, distance estimation contributes to related tasks such as object localization and identification, helping users better understand spatial arrangements within a scene.

\subsection{Temporal Relationships}

The ability to analyze temporal changes is particularly important when distinguishing between static and dynamic objects, a fundamental characteristic of point cloud data.

We categorize visual tasks related to temporal relationships into two main groups: {\em shape change} and {\em tracking}. Although these tasks focus on different aspects of temporal analysis, they are fundamentally extensions of the spatial and parts-to-whole relationships discussed earlier. By incorporating time as an additional dimension, they build upon tasks such as object detection, segmentation, and spatial positioning, extending them across multiple frames.

{\bf Shape change.} While object tracking focuses on movement, another crucial visual task is understanding how objects themselves change in shape and structure over time. This involves analyzing transformations in an object's geometry, surface characteristics, or overall morphology across different time steps. A common example is forest growth monitoring, where point clouds are used to track the emergence of new trees, branches and overall canopy over time \cite{golla_temporal_2020}. Similarly, in environmental studies, point clouds help assess changes in the morphology of a landscape \cite{Tateosian_Mitasova_Thakur_Hardin_Russ_Blundell_2014}. Shape change analysis can occur at different scales, ranging from small-scale structural modifications, such as detecting cracks in infrastructure or identifying deformations in mechanical components, to large-scale transformations, such as monitoring urban expansion or tracking long-term environmental shifts \cite{golla_temporal_2020}.

{\bf Tracking.} refers to the task of following the movement of points or objects across multiple frames in a point cloud time series. Given the position of objects in an initial frame, the goal is to determine their locations in subsequent frames \cite{Guo2021}. This task is essential in applications, such as autonomous driving, where vehicles and pedestrians must be tracked over time \cite{vaquero_deconvolutional_2017}.

Depending on how dynamic point clouds are captured, object tracking can present different challenges. If objects are moving while the point cloud is being recorded, their shape may appear distorted which can make object tracking more complex, as motion artifacts may obscure clear object boundaries \cite{Pomerleau_Liu_Colas_Siegwart_2012}. If dynamic events are captured in discrete frames, the task is to identify the same object across different frames while accounting for potential occlusions, missing data, or varying point densities \cite{Pomerleau_Liu_Colas_Siegwart_2012}. A related task is trajectory estimation, which goes beyond tracking object positions to infer the path and movement characteristics of objects. This includes determining both direction and speed, which are crucial for applications such as robotics and traffic analysis \cite{Guo2021}.

\section{Design choices (How)}
\label{sec:design}

Now that we have outlined the types of data and user tasks, we turn our attention to organizing the existing design techniques used for visualizing point clouds. As with previous sections, the aim here is not to provide an exhaustive review of the technical challenges in point cloud visualization, such as data structures, hardware requirements, or specific rendering operations. Instead, our focus is on the design choices explored by the research community. While some technical aspects may be touched upon, they will be discussed only in relation to the design solutions they enable.

Our classification is organized starting from the type of primitives that are being rendered and builds down the different types of techniques that we have seen applied to change the appearance of such primitives and convey different types of information about the data. As a result, we structure this section in three parts organized techniques as {\em point-based}, {\em grid-based}, and {\em geometry-based}. An overview of the proposed task taxonomy is presented in Table \ref{tab:designs} where we provide a categorization of the identified design choices and a connection with the supported user tasks and data types.

Two broader design choices are not explicitly categorized in this table, as they are common across all rendering primitives and inherited from general 3D visualization approaches.

{\bf 2D vs. 3D projections.} The debate between 2D and 3D representations is longstanding in the visualization community, with studies yielding mixed results over the years. This question extends to point clouds, where choosing between 2D projections or full 3D renderings remains an area of active study \cite{Aygar_Ware_Rogers_2018,gruchalla_structure_2021}. 2D projections can simplify the amount of data shown by collapsing depth information into a single plane. This reduces occlusions and can make patterns more easily recognizable. However, this reduction in complexity comes at the cost of spatial accuracy, as depth relationships and finer geometric details may be lost. In contrast, 3D renderings preserve the full spatial structure, allowing for precise spatial reasoning and depth perception. Yet, they introduce challenges such as occlusion and perspective distortions.

{\bf Realism vs. data-centric visualization.} A substantial body of work explores the use of color \cite{Goude2021}, shading \cite{alexa_computing_2003}, and lighting \cite{sabbadin_high_2019} to enhance realism in point cloud rendering. However, the impact of realistic rendering on low-level analytical tasks compared to purely data-centric visualizations remains largely unexplored.

In our review, we found that this aspect has received little attention in the literature. While realism can improve immersion and familiarity in certain applications, data-centric visualizations are often designed to reduce cognitive load by emphasizing structures, spatial relationships, or statistical properties directly relevant to analysis. The extent to which realistic rendering influences task performance remains an open question in the field.

\begin{table*}[ht]
\centering
\caption{Overview of rendering techniques for spatial point cloud visualization, categorized by rendering primitives. For each technique, we highlight the ideal data types and tasks it supports, along with representative papers that have applied the technique for these purposes. It is important to note that these associations are based on empirical observations and our judgment, as we found no formal evaluations proving one technique to be quantitatively superior to another.}
\label{tab:designs}
\resizebox{\textwidth}{!}{
    \begin{tabular}{llllrrr}
    \toprule
     Design Choices & Category & Sub-Category &   Technique &   Data Types(What) &  Tasks (Why) &  Papers \\
    \midrule
    \multirow{8}{*}{Point-based} 
        &   \multirow{4}{*}{Sampling} &  \multirow{2}{*}{Filtering} &    Data Based &         object &    \taskparts{part-to-whole} &    \cite{schutz_fast_2020,de_silva_edirimuni_straightpcf_2024} \\
        &         &         &    View Based &          object, scene, multi-view &    \taskparts{part-to-whole} &     \cite{katz_direct_2007, gruchalla_structure_2021} \\
        \cmidrule(l){3-7}
        &         & \multirow{2}{*}{Resampling} &    Geometries &     object, scene, single view &    \taskparts{part-to-whole} &     \cite{li_anisotropic_2010,shengjun_liu_iterative_2012}    \\
        &         &         &      Learning &   scene, single view &    \taskparts{part-to-whole} & \cite{sung_data-driven_2015,xiao_lgsurnet_2024} \\
        \cmidrule(l){2-7}
        & \multirow{5}{*}{Attributes} &        &           Hue &     * &    \taskparts{part-to-whole}, \taskspatial{spatial} &  \cite{burt_extracting_2019,Yu2019}  \\
        &        &        &    Saturation &      * &    \taskspatial{spatial} &   \cite{Musialik2015a,wang_lidar_2018} \\
        &        &        &  Opacity &      scene, dynamic  &    \taskspatial{spatial}, \tasktemporal{temporal} &    \cite{seemann_soft_2018,Uchida_2020} \\
        &        &        &       Symbols &     scene, dynamic &    \taskspatial{spatial}, \tasktemporal{temporal} &    \cite{botsch_high-quality_2005,Wegen2024} \\
        &        &        &          Size &        scene &    \taskspatial{spatial} &    \cite{Hanula2015,seemann_soft_2018}  \\
\midrule
    \multirow{4}{*}{Grid-based}
        &        &        &     Aggregate &     object, scene &    \taskparts{part-to-whole}, \taskspatial{spatial} &    \cite{Elfes1989,bock_visualizationbased_2017} \\
        &        &       & Agg. + Derive &     particle &    \taskparts{part-to-whole}, \taskspatial{spatial} &     \cite{Shivashankar2016a,bock_visualizationbased_2017}  \\
        &        &        &       Uniform &     * &    * &       \cite{rusu_3d_2011}   \\
        &        &        &      Adaptive &         * &    * &      \cite{schutz_fast_2020}   \\
\midrule
    \multirow{5}{*}{Geometries-based}
        &    Annotations    &       &       &    object, scene, dynamic &    * &   \cite{Zeng2018,Guo2021} \\
        \cmidrule(l){2-7}
        &   \multirow{2}{*}{Lines}     &       &      Outlines &      object, single view &    \taskparts{part-to-whole} &    \cite{pauly_multiscale_2003,wang_fitting_2006} \\
        &        &        &      Skeleton &       object, dynamic &    \taskparts{part-to-whole}, \tasktemporal{temporal} &    \cite{huang_l1_2013,yin_morfit_2014} \\
        \cmidrule(l){2-7}
        &   \multirow{2}{*}{Surfaces}     &    & 2D Primitives &     scene &    \taskparts{part-to-whole}, \taskspatial{spatial} &     \cite{Poux2022,Raffo2022}   \\
        &        &        &        Meshes &      object &    \taskparts{part-to-whole}, \taskspatial{spatial} &   \cite{poullis_photorealistic_2009, wang_neural-singular-hessian_2023} \\
    \bottomrule
    \end{tabular}}
\end{table*}

% \begin{table*}[ht]
% \centering
% \caption{Overview of rendering techniques for spatial point cloud visualization, organized by rendering primitive and annotated with representative citation numbers.}
% \label{tab:designs}
% \begin{tabular}{@{}llp{8cm}@{}}
% \toprule
% \textbf{Category} & \textbf{Technique} & \textbf{Representative Works} \\
% \midrule
% \multirow{8}{*}{Point-based} 
%     & Data-based Filtering & \cite{} [3], [15], [35], [39], [76], [10] \\
%     & View-based Filtering & [22], [25], [30], [40], [58] \\
%     & Splat-based Filtering & [16], [18], [49], [56], [68] \\
%     & Resampling (Geometric) & [10], [39], [35] \\
%     & Resampling (Learning-based) & [69], [41], [53], [60] \\
%     & Attribute Visualization & [68], [74], [19] \\
%     & Curvature Estimation & [33], [66], [46], [23] \\
%     & Normal Estimation & [14], [52], [34], [55], [70], [38], [66] \\
% \midrule
% \multirow{2}{*}{Grid-based} 
%     & Voxelization / Grid Structuring & [4], [37] \\
%     & Level-of-Detail (LOD) & [57] \\
% \midrule
% \multirow{4}{*}{Geometry-based} 
%     & Boundary / Line Representations & [65] \\
%     & Skeletonization & [26], [71], [13] \\
%     & Primitive Fitting / Segmentation & [47], [51], [32], [36], [48], [73] \\
%     & Mesh Reconstruction & [7], [12], [27] \\
%     & LOD for Meshes & [57] \\
% \bottomrule
% \end{tabular}
% \end{table*}

\section{Point-based designs}
\label{sec:points}

One approach to visualizing point clouds is to render them directly as individual entities in the scne. This section describes the two key design choices we have encountered when reviwing point-based visualization papers, how many points should be visualized, and how to visualize the point attributes.

\subsection{Sampling techniques}

This section describes sampling design choices that reshape the distribution of points by selectively removing, adding, or repositioning them. These techniques are categorized into two main approaches: filtering, which reduces the dataset by removing points, and resampling, which enhances or redistributes points.

\subsubsection{Filtering}
\label{sec:filtering}
Point cloud filtering techniques focus on preprocessing and selecting relevant subsets of points to visualize to improve clarity and reduce visual clutter. We classify these techniques as either data based or view based.

{\bf Data based filtering.} Data-based filtering improves point cloud quality by selectively removing undesired points based on structural properties or point attributes. Early research in this area aimed at reducing clutter by enforcing a uniform point distribution across the point cloud \cite{oztireli_analysis_2012,schutz_fast_2020}. However, these geometric techniques imposed a fixed resolution across the entire point cloud, ignoring local surface characteristics and non-spatial features like color and light intensity. 

To address these limitations, adaptive filtering techniques have been proposed, applying different filtrations to different regions of the mesh. For example, patch-based denoising techniques identify local neighborhoods around each point and construct patches based on geometric alignment \cite{rosman_patchcollaborative_2013}. By applying different filters to different regions of the point cloud, these methods effectively reduce noise while preserving sharp geometric features.

More recently, deep learning-based filtering has gained attention for its ability to learn adaptive sampling strategies. These approaches refine point cloud data by denoising, enhancing geometric features, and preserving sharp details for improved surface reconstruction \cite{xu_globally_2023, de_silva_edirimuni_straightpcf_2024}. Deep learning methods offer greater flexibility than traditional algorithms by dynamically adjusting filtering strategies based on learned patterns, making them particularly effective for complex, high-resolution point clouds.

Overall, data-based techniques help reduce clutter and noise while strategically concentrating points where they are most needed, particularly around sharp features and key characteristics of objects.

{\bf View based filtering.}
The hidden-surface problem \cite{Sutherland1974}, determining which parts of a scene should be visible in a rendering, is a fundamental challenge in computer graphics. In point cloud visualization, view-dependent filtering selectively displays points based on the user viewpoint. This technique provides computational benefits for real-time rendering by reducing the rendering workload \cite{schutz_progressive_2020,hofsetz_objectspace_2008}, but importantly, it also enhances viewer perception by prioritizing visually meaningful points and reducing visual clutter \cite{metzer_z2p_2022}.

A key technique for view-based filtering is the Hidden Point Removal (HPR) operator introduced by Katz et al. \cite{katz_direct_2007}. While effective for isolated objects, this method faces scalability issues in large scenes and real-time scenarios because of expensive convex hull computations \cite{machado_e_silva_image_2014}. To address these limitations, progressive refinement techniques were proposed, significantly enhancing real-time updates and scalability \cite{hofsetz_objectspace_2008,schutz_progressive_2020}.

% Hofsetz et al. \cite{hofsetz_objectspace_2008} introduced an efficient object-space method for visibility ordering in point-based rendering, pre-sorting primitives to rapidly handle viewpoint changes. In contrast, Guennebaud et al. \cite{guennebaud_dynamic_2008} presented an adaptive upsampling approach that dynamically refines point clouds using view-dependent error metrics, improving rendering quality by increasing sampling density near visually critical features like silhouettes and high-curvature regions. Unlike Hofsetz et al.'s explicit depth sorting, Guennebaud et al. integrate visibility directly into splatting, enhancing visual continuity.

Splat-based visualization techniques inherently incorporate view-dependent filtering, improving rendering efficiency and reducing perceptual artifacts. These techniques represent points as disc-like primitives, balancing computational efficiency with visual quality \cite{wu_optimized_2004,ren_object_2002}. A notable advancement in this direction is 3D Gaussian Splatting, which optimizes per-point Gaussian parameters to approximate a continuous surface better, reducing visual artifacts and enhancing perceptual quality \cite{kerbl3Dgaussians}. While blending point-based and surface-based rendering characteristics, we categorize it under point-based approaches because each Gaussian fundamentally represents discrete points, each optimized in shape, size, and opacity for improved viewer perception.

\subsubsection{Resampling}

%In recent years the attention has shifted to using deep learning networks to learn adaptive sampling strategies.

%A few approaches have focused on upsampling techniques to increase the resolution of point clouds. The typical approach is to train the model using a large set of shapes for which the ground truth surface is known and then generalize the results to point clouds with unseen shapes \cite{xiao_lgsurnet_2024}.

%Other approaches instead are able to improve the quality of the input point cloud directly without external training data by iteratively working on small patches of the input point cloud \cite{metzer_self-sampling_2022}.

Resampling techniques modify the spatial distribution of points, either by upsampling to increase resolution or by moving existing points to improve geometric fidelity or uniformity. These methods are distinct from filtering, as they do not discard points but rather reorganize or generate new ones to enhance visualization quality.

{\bf Geometric Resampling.}
Early approaches aimed at achieving a more uniform or perceptually meaningful distribution of points. For instance, methods like bilateral blue noise sampling \cite{chen_bilateral_2013} move points to enforce an even yet feature-aware spacing, placing more points near edges while keeping smooth regions sparser. Similarly, uniformization strategies \cite{luo_uniformization_2018} and anisotropic sampling methods \cite{li_anisotropic_2010} adjust point positions to improve surface representation without generating new points.
In contrast, approaches like {\em iterative consolidation} \cite{shengjun_liu_iterative_2012} takes a different approach by directly combining point movement with upsampling. It uses a step-by-step process that spreads points out using repulsion forces and then adjusts them to better fit the surface. Although not learning based, it is a good example of how early methods worked to clean up and enhance sparse or noisy point clouds by making the point layout denser and more accurate.

{\bf Learning-Based Resampling.}
More recently, deep learning approaches have been proposed to either upsample point clouds or consolidate noisy input by moving points. LGSur-Net \cite{xiao_lgsurnet_2024} is a deep network that learns to upsample sparse point clouds by inferring missing geometry from training data. In contrast, self-supervised methods like Self-Sampling \cite{metzer_self-sampling_2021} refine the input point cloud without external supervision, by iteratively relocating points to reduce reconstruction loss across small patches. Other approaches like PointProNets \cite{roveri_pointpronets_2018} improve point distributions by learning local surface structures and projecting points accordingly. Rather than discarding or adding points arbitrarily, these methods learn to relocate and upsample points in a structured manner, allowing to reconstruct missing shapes and objects.

Shape completion methods aim to restore missing regions in point clouds by synthesizing or repositioning points, contributing directly to the perceptual coherence of the shape. Earlier approaches \cite{sung_data-driven_2015} introduced structural priors learned from databases of segmented shapes, using symmetry detection and part-based reasoning to infer plausible completions even under severe occlusion. More recent techniques\cite{miao_end--end_2021} employ end-to-end networks with encoder–decoder architectures that capture both global shape and local geometric features, enabling the generation of dense, uniformly distributed completions. 

\subsubsection{Relations with data types and tasks}

Sampling techniques are typically used for static point clouds, both single objects and scenes. They seem particularly well suited for parts-to-whole tasks since they typically minimize clutter and emphasizing relevant structures (see Table \ref{tab:designs}). Resampling techniques are often proposed for single-view data types since they can effectively fix missing data and provide shape completion capabilities \cite{sung_data-driven_2015}. Despite the relatively well-recognized applications, there is a surprising lack of empirical research investigating how different sampling strategies influence user tasks.

A key exception is the work of Gruchalla et al. \cite{gruchalla_structure_2021}, which provides valuable insight into human perception of structure within noisy point clouds. Their study examined whether users could recognize familiar 3D objects under various noise rates. With 128 participants, they tested various conditions, including 3D scatterplots with rotation, xy-translation, z-translation, and static 2D projections. Their results showed that rotation significantly enhanced object recognition, while static projections and translations were largely ineffective. Remarkably, participants were able to identify objects even in point clouds with exceptionally high noise levels, far beyond what a machine could tolerate for downstream tasks. This suggests that human perception is highly resilient to noise, challenging assumptions about the necessity of aggressive filtering. However, this study focused exclusively on object identification in single object datasets, leaving broader questions about the impact of sampling techniques on other analytical tasks unanswered. 

% For example,in tasks involving tracking moving objects in dynamic point clouds, does filtering improve users’ ability to maintain object continuity across frames? Similarly, in segmentation tasks, does reducing visual noise enhance the accuracy of detected shapes? These open questions highlight the need for further research into the perceptual and cognitive effects of sampling, ensuring that these techniques are not only computationally efficient but also beneficial for human-data interaction.

\subsection{Visualization of attributes}

\begin{figure}[h]
    \centering
    \includegraphics[width=0.95\linewidth]{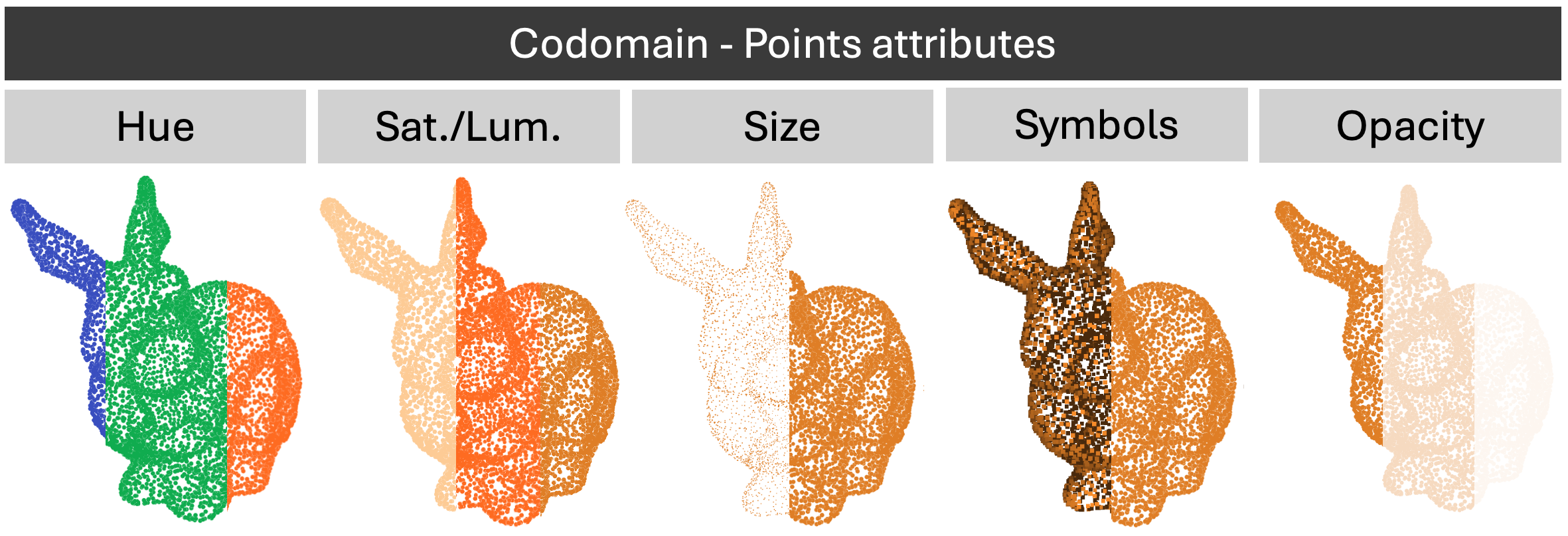} % Replace with your image filename
    \caption{Visual encodings that can be used for modifying the appearance of points based on codomain attributes.}
    \label{fig:codomain}
\end{figure}

Regarding the visualization of quantitative and categorical attributes we found much more similarity with the visual channels typically employed in abstract data visualization (see Figure \ref{fig:codomain}).

Quantitative attributes in point clouds are typically visualized through color mapping that involve saturation or luminance. This is a very common approach but prone to the well-known challenges in selecting appropriate and effective color maps \cite{Zhou2016}. Opacity can also be used to emphasize or de-emphasize points not only based on spatial or view based rules \cite{seemann_soft_2018} but also based on their attribute values \cite{Musialik2015a}. For instance, points with lower confidence scores or lower intensity in LiDAR scans can be rendered more transparently, directing attention to more significant structures. Finally, the size channel can be used to change the size of the represented points given than this can introduce further cluttering which is already a common problem in point cloud rendering.

Qualitative attributes, can be mapped using categorical color maps. Assigning distinct colors to different categories. Differently from abstract data is much more rare, but not impossible, to encode these attributes through symbolic variations such as different glyph shapes. This type of encoding is mostly employed in splat-based rendering techniques \cite{botsch_high-quality_2005}.

A special case is dedicate to color attributes that represent the color of the object sampled by the point. In this case the single RGB components are blended together in an effort to represent realistic scenarios. Recently methods have been proposed that use deep neural networks to generate color attributes based on new viewpoints or to transfer color stiles from images to 3D point clouds \cite{Goude2021}.

On top of attributes provided with the point cloud we should remember point clouds lack inherent structure and shape while representing structured objects. For this reason computing (and visualizing) derived information such as curvature, normal orientation, and other differential properties is essential to enhance their usability.

{\bf Curvature and geometric features.} Once computed, curvature is a scalar value that can be plotted like any other quantitative attribute.

Researchers have approached curvature estimation through geometric methods that rely on local approximations of surface patches. These methods typically compute curvature using techniques such as least-squares fitting or PCA-based analysis of local neighborhoods \cite{lejemble_stable_2021}. While effective under ideal conditions, these techniques are sensitive to noise and sampling irregularities.

To address this issue, more recent efforts have explored implicit neural representations as a more flexible alternative. In this line of work, curvature is not computed directly from local geometry but inferred from continuous surface fields that capture the underlying shape. These methods adapt more easily to variations in point density and preserve high-frequency geometric details better than traditional estimators \cite{wang_neural-singular-hessian_2023}.

In parallel, researchers have also focused on feature detection techniques that identify high-curvature structures without explicitly computing curvatures. For example, multi-scale surface variations can be use used to differentiate between smooth and sharp regions by analyzing local neighborhoods at varying scales  \cite{pauly_multiscale_2003}.

{\bf Normals.} 
Just like curvature, normals are a fundamental piece of information for point cloud processing \cite{boulch_deep_2016}. A widely used approach is to compute normals through local surface approximations, typically by fitting planes to small neighborhoods via Principal Component Analysis (PCA) \cite{10.1145/142920.134011}. These methods are conceptually simple but remain sensitive to noise and variations in point density.

To improve robustness, researchers have developed enhanced local techniques such as RANSAC-based estimators  \cite{li_robust_2010}, or Voronoi-based constructions \cite{rosenthal_enclosing_2009}. However, these approaches still estimate normals independently at each point, often resulting in inconsistent orientations. 

To address these limitations, a number of global optimization strategies have been proposed. These methods frame normal orientation as a consistency problem across the entire point cloud. For example, some techniques model this task as a graph-based energy minimization \cite{schertler_towards_2017}. Others adopt a field-regularization approach, such as the Regularized Winding Field (RWF) method \cite{xu_globally_2023}. While these optimization-based methods have substantially improved normal estimation, their evaluation has largely focused on computational accuracy, with limited understanding of their perceptual impact on visualization tasks.

\subsubsection{Relation with data types and tasks}

We found that color is widely used across all data types, and are particularly recurring for part-to-whole and spatial tasks (see Table \ref{tab:designs}). Categorical colormaps are commonly applied to support part-to-whole tasks by distinguishing different regions or components \cite{burt_extracting_2019}, while sequential colormaps are often used to enhance spatial perception \cite{Musialik2015a}.

Opacity and symbols, on the other hand, are frequently employed in complex scenes or dynamic contexts \cite{seemann_soft_2018,botsch_high-quality_2005}. Notably, these are the first visual channels in our taxonomy that we found specifically applied to dynamic datasets for temporal tasks.

Although curvature and normal estimation are not included in Table \ref{tab:designs}, as they are not direct design choices for visual encoding, the decision to compute and visualize them introduces important trade-offs that warrant further study. Estimating these values requires balancing precision, computational efficiency, and the perceptual impact on the final visualization. A more systematic investigation is needed to determine the optimal methods and configurations for computing these features in real-world applications.

\section{Grid-based designs}
\label{sec:grid}

Grid-based rendering organizes 3D spatial data into a structured grid by partitioning space into discrete cells. This structured approach simplifies the processing and visualization of large point clouds by converting unstructured data into a more manageable format.

{\bf Aggregate and Derive.} One fundamental design choice in grid-based methods is whether to perform simple aggregation or aggregation with derived values. In simple aggregation, multiple points are merged into a single grid cell, typically by averaging or selecting representative values. This approach is computationally efficient and reduces memory requirements while maintaining a coarse approximation of the original data. However, it may lead to a loss of fine-grained details that could be important for downstream analysis.

Aggregation with derived values, instead, enhances the stored information by computing new attributes based on the original point cloud data. These derived attributes can encode additional semantic or geometric properties, improving both visualization and analytical capabilities. Common examples include occupancy probabilities \cite{Elfes1989}, which indicate whether a cell is likely to contain a surface, point densities \cite{Ladicky2017}, which provide insight into spatial distribution and clustering, and signed distance functions (SDFs) \cite{Park2019}, which encode distance to the nearest surface.

{\bf Uniform or Adaptive.} To improve scalability and performance, Level of Detail (LOD) models are often employed \cite{schutz_gpuaccelerated_2023}. While these models are not only important for grid-based approaches, they can directly affect the voxelization process by imposing a non-uniform subdivision of the space. Hierarchical data structures like octrees \cite{Scheiblauer2011} and kd-trees \cite{Goswami2013} are often used to dynamically adjust the grid resolution based on spatial complexity or viewing parameters. This adaptive approach helps optimize rendering efficiency by maintaining detail where needed while reducing computational overhead in less critical regions. Many widely used tools such as Open3D, PCL, and Potree provide built-in voxelization techniques to facilitate efficient point cloud management and visualization \cite{rusu_3d_2011,schutz_fast_2020}.

% Efficient Level of Detail (LOD) construction is a key factor especially for handling large-scale point clouds in real-time applications. Traditional CPU-based LOD methods are computationally expensive, limiting interactivity. Schütz et al. \cite{schutz_gpu-accelerated_2023} address this by introducing a GPU-accelerated LOD generation technique that processes up to 4 billion points per second, achieving an 80× to 400× speedup over CPU methods. Their hybrid voxel-point octree structure reduces memory consumption by voxelizing lower LODs, enhancing streaming efficiency while maintaining high visual fidelity through color filtering. Though currently limited to in-core processing, this approach represents a significant step toward scalable, high-performance point cloud visualization.

\subsection{Relation with data types and tasks}

Grid-based rendering provides an effective balance between performance and detail, making it well-suited for tasks requiring parts-to-whole and spatial understanding. Simple aggregation helps reduce noise and computational load \cite{Scheiblauer2011}, while aggregation with derived attributes enhances tasks like segmentation by incorporating richer geometric properties. These design choices are particularly well-suited for particle-based datasets, where aggregation and derivation (e.g., density estimation) help analyze spatial distributions \cite{Shivashankar2016a}.

Intuitively, the choice between uniform and adaptive grids impacts task effectiveness. Uniform grids offer predictable structure, ideal for volumetric rendering and feature extraction. On the down side, fixed grid sizes may lead to oversampling or undersampling, and lower resolutions can introduce blocky artifacts or loss of fine details \cite{ran_learning_2021}. We say intuitively  because we have not found empirical studies focusing on this aspect of the visualization design choices.

% \todo[inline]{We had the following citation and text in the document but this does not seem related to point clouds at all. We have to be careful of this type of mistakes because this can easily kill our paper.}

% This technique also facilitates spatial analysis by enabling tasks like noise filtering, downsampling, and efficient data querying. Grid-based rendering strikes a balance between performance and detail, making it a valuable tool for managing complex point clouds in real time\cite{lin2023high}. However, it comes with notable limitations. Fine details can be lost, memory usage can be high, and blocky artifacts may happen, especially at lower resolutions. Fixed grid sizes often struggle with varying point densities, leading to a loss of precision in the original data. While it’s great for simplified overviews or general analysis, it’s less ideal for high-detail visualizations or real-time applications where accuracy and responsiveness are crucial\cite{lin2023high}.

\section{Geometry-based designs}
\label{sec:geometry}

\begin{figure}[h]
    \centering
    \includegraphics[width=0.95\linewidth]{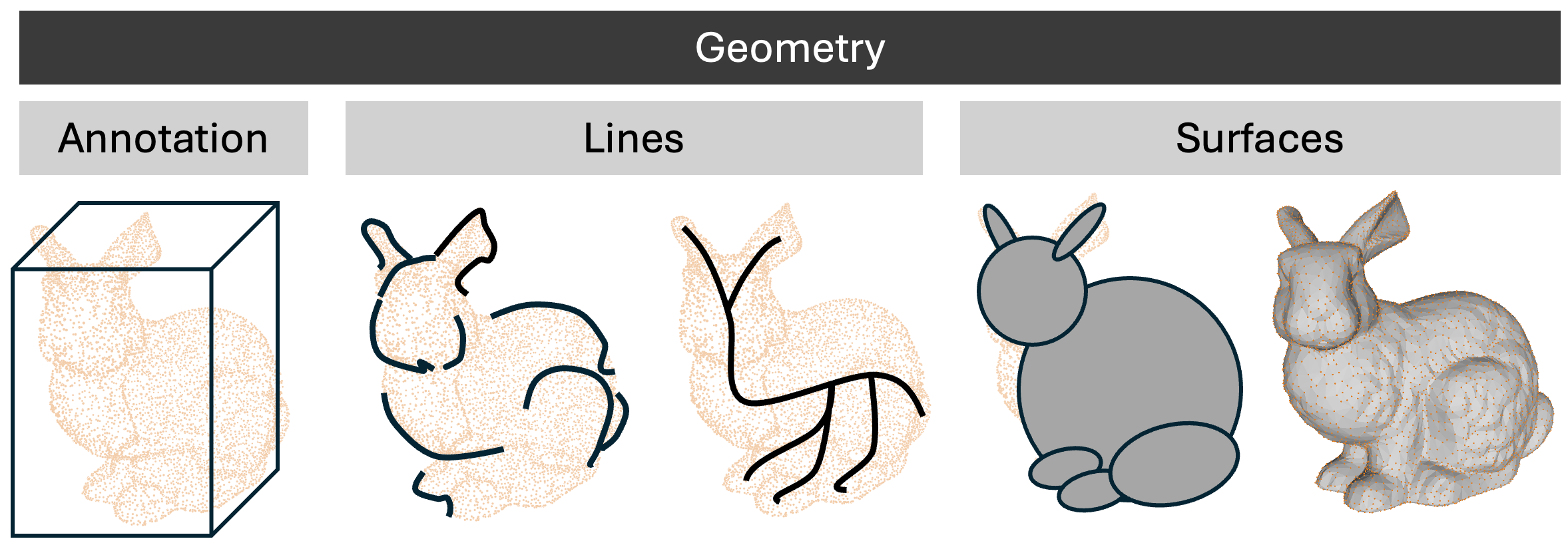} % Replace with your image filename
    \caption{Geometry based techniques organized in three subcategories, annotations, lines, and surfaces.}
    \label{fig:geometry}
\end{figure}

Beyond point-based and grid-based rendering common ways to visualize point clouds are geometry-based techniques, explicitly representing the underlying geometric structure of a point cloud (see Figure \ref{fig:geometry}). These techniques serve multiple purposes. {\em Annotation-based} methods use simple geometric primitives to highlight objects. {\em Line-based} methods emphasize object boundaries, feature edges, or connectivity patterns between points. {\em Surface-based} approaches, aim to generate continuous surfaces from discrete points.

\subsection{Annotations} \label{sec:skeleton}

Annotations are widely used to enhance the interpretability of point cloud visualizations by explicitly marking areas of interest. Common annotation techniques include points, lines, and bounding boxes, which highlight specific structures, patterns, or objects within the data. Point annotations are often used to tag individual locations, providing labels or metadata for specific features, such as landmarks or ground truth labels in machine learning datasets \cite{Guo2021}. Lines help emphasize relationships between different regions, making them useful for structural analysis (e.g., trajectory visualization). Bounding boxes, whether axis-aligned or oriented, are widely used for object detection and segmentation.

Annotations are particularly valuable in conjunction with machine learning and deep learning applications, where they serve as a bridge between automated methods and human interpretation \cite{Guo2021}. They provide visual explanations of model outputs, allowing users to verify predictions, highlight detected objects, or analyze misclassifications. 

\subsection{Curves and Lines} \label{sec:skeleton}
Line-based representations are widely used to enhance the visualization and structural understanding of point clouds, aiding in shape perception, editing, and domain-specific applications. These can be either {\em sketch lines} or {\em skeletons}.

{\bf Sketch lines.} Several techniques focus on feature-line enhancement to improve the visibility of object boundaries and structural outlines. These line-based representations are particularly effective in aiding scene comprehension, especially in cluttered or sparse point clouds. Feature lines can be extracted in either object space, using neighborhood geometry, or in image space, leveraging depth or normal buffers. Some methods generate explicit geometric lines (e.g., triangle strips or splines), and can be used for either static as well as dynamic data \cite{wang_fitting_2006,chandler_illustrative_2013}.

Many of these approaches rely on curvature-based or surface-aware methods to identify and emphasize salient features. For example, Pauly et al. \cite{pauly_multiscale_2003} compute feature lines based on curvature estimates, enabling the creation of line drawings that enhance the perception of shape in point-sampled surfaces. Similarly, more recent techniques use graph-based or statistical descriptors, such as structured local binary patterns (SGLBP), to detect discontinuities efficiently while maintaining robustness to noise \cite{guo_sglbp_2022}.

Feature-line extraction has also been explored in domain-specific applications. For instance, Poullis and You (2009) \cite{wang_lidar_2018} developed a method for extracting building outlines from LiDAR point clouds. Their approach converts 3D LiDAR data into 2D regular grids, applying segmentation techniques to detect roof contours. These contours, often visualized as linear boundary representations, provide a structural overview of urban environments and serve as an intermediate step before full mesh generation.

{\bf Skeletons.} 
Skeleton extraction from point clouds generates simplified 1D structures that support shape abstraction, editing, and visualization. Traditional techniques typically rely on noise-free data and accurate surface normals  \cite{tagliasacchi_curve_2009}, limiting their applicability to raw or incomplete scans. More recent developments relax these constraints through robust estimators and interaction-driven pipelines. One class of methods uses robust statistical techniques to infer skeletal structures directly from noisy and unorganized data, avoiding surface reconstruction altogether. These approaches often leverage localized computations, such as L1-medians, to iteratively converge toward stable skeletal representations \cite{huang_l1_2013}. Another category involves user-guided reconstruction, where interactive frameworks offer tools to edit and refine skeleton topologies for improved geometric control. Such methods allow users to address sparsity or missing regions manually, making them effective for complex or incomplete data \cite{yin_morfit_2014}. Skeletonization also plays a critical role in domain-specific modeling, such as representing tree-like structures in ecological or urban scenes \cite{cardenas_modeling_2023}.

\subsection{Surfaces and meshes}
Reconstructing a high-quality surface from a set of disconnected 3D points is a challenge that comes with big rewards. Unlike direct point-based rendering, mesh-based techniques provide continuous surface representations, enhancing the perception of object boundaries, spatial relationships, and geometric details. This class of the methods is organized into two groups: {\em 2D primitives}, and {\em meshes}.

{\bf 2D Primitives.}
These methods focus on identifying simple shapes, such as planes, cylinders, or spheres, that approximate regions within a point cloud. Many of these methods are tailored to specific domain applications, optimizing feature extraction based on the characteristics of the targeted environment. For instance, techniques designed for single CAD models differ significantly from those used in large-scale outdoor scenes or indoor architectural scans \cite{Poux2022, Romanengo2024}.

The computation of geometric primitives is often closely linked to point cloud segmentation. Segmentation divides the point cloud into regions with similar characteristics, allowing primitives to be fitted more effectively. Various approaches have been employed for this purpose, including k-means clustering \cite{Kong2014}, region-growing methods \cite{Li2020}, and the Hough transform \cite{Raffo2022}. 

In recent years, machine learning has become increasingly prominent in this field, particularly when large amounts of labeled training data are available. Deep learning approaches have demonstrated strong performance in segmenting point clouds and identifying underlying geometric structures, outperforming traditional methods in complex scenarios with high variability \cite{Armeni2016a,Zhao2018}. 

{\bf Meshes.} A comprehensive survey by Berger et al. \cite{Berger2017} highlights the wide range of strategies for surface reconstruction, from classical interpolation methods to modern learning-based approaches. Our review of recent literature confirms that mesh reconstruction from point clouds remains an open problem. New graph-based techniques introduce topological constraints to improve mesh generation, reducing common artifacts in traditional methods while preserving fine details \cite{cui_surface_2024}. 

As in other point cloud representations, deep learning combined with mathematical optimization has shown promise in improving mesh generation \cite{huang_neural_2022}. These approaches aim to refine surface quality by learning complex patterns in point distributions, operating directly on the point cloud, its gridded counterpart, or both \cite{Fan2024}. Unlike classical methods that rely on explicit normal or curvature estimation, deep learning-based techniques can infer an implicit surface directly from raw point data \cite{lin_surface_2023, wang_neural-singular-hessian_2023}. This shift enables more flexible and robust reconstructions, reducing dependency on precomputed geometric properties.

\subsubsection{Relation with data types and tasks}

The primary role of geometry-based approaches is to make implicit structures explicit. This is a powerful tool to reduce the number of points visualized by constructing explicit surfaces to control visibility. While raw point clouds provide a direct representation of 3D spatial information, it is natural to hypothesize that their lack of connectivity limits interpretability, especially for tasks that require understanding spatial relationships, tracking changes over time, or extracting meaningful structures. While this is a natural hypothesis to produce we have found no empirical studies comparing raw point clouds with meshes for fulfilling various tasks. Qualitatively we know that dense point clouds contain a high number of points that can be visually overwhelming and difficult to interpret.

To this end, annotation techniques help tremendously for all kind of tasks since they introduce specific annotations (e.g., points, lines, polygons) tailored to the application of choice \cite{Armeni2016a,vaquero_deconvolutional_2017}.

Techniques such as boundary extraction and surface meshing help define object contours, making it particularly easy to recognize shapes and spatial separations. Skeletonization and curve-based methods provide a more intuitive representation that works particularly well in dynamic scenarios \cite{Tateosian_Mitasova_Thakur_Hardin_Russ_Blundell_2014}. 

In datasets where point clouds are enriched with additional attributes, such as scientific measurements, simulation outputs, or material properties, geometry-based approaches also serve a crucial function. Attributes such as temperature, velocity, or material type can be challenging to interpret when only represented as individual points. Structured visualization techniques, such as polygonal approximations and meshing, allow for smoother interpolation of these attributes across surfaces.

\section{Discussion}
\label{sec:discussion}

Our review activity has provided insights into the effectiveness, limitations, and areas for future exploration in point cloud visualization. By adopting a task-oriented perspective, we have identified several discussion points.

{\bf Abstract data visualization.} A natural comparison can be drawn between point cloud visualization and scatterplot-based techniques for abstract data \cite{Sarikaya2018}. We think the most crucial distinction, however, is that in abstract data visualization, each point is treated as an independent object. In contrast, spatial point clouds represent objects as collections of points, which fundamentally changes the nature of the data and its visualization. This shift alters the effectiveness of various encoding strategies. Techniques that work well for visualizing individual data points in scatterplots may not translate directly to point cloud visualization, where spatial relationships, depth, and structure play a different role. As a result, existing empirical evaluations of scatterplot techniques are not easily applicable to point cloud data but, similar empirical evaluations for spatial point clouds remain scarce.

{\bf Machine learning.} When comparing our taxonomy with machine learning-based taxonomies for point cloud processing, we observe intriguing parallels and gaps \cite{Guo2021}. Both taxonomies categorize techniques based on how they process and extract information from point clouds. However, our analysis has uncovered underexplored areas in visualization research, particularly regarding the partitioning of point clouds into components. In machine learning, models often divide point clouds into smaller, semantically relevant regions to improve feature extraction and facilitate downstream tasks. Surprisingly, this approach has not been adopted in visualization. We hypothesize that incorporating this strategy into visualization methods could enhance users' ability to navigate and interpret large-scale 3D data, especially for tasks that involve part-to-whole relationships.

{\bf Task-oriented evaluations.} A major challenge in point cloud visualization is the lack of standardized evaluation metrics for measuring the effectiveness of different techniques. Many existing methods either aim to enhance high-level tasks like scene understanding \cite{wu_physical_2024}, or rely on subjective user preferences for evaluation \cite{dumic_point_2021}, or focus on domain-specific objectives that are difficult to generalize across applications \cite{bock_visualizationbased_2017}. This inconsistency makes it difficult to assess which techniques are most effective for low-level analytical tasks.
Our proposed taxonomy and task categorization seek to address this gap by offering a structured framework for evaluating visualization approaches. By explicitly defining user tasks and objectives, we aim to support the development of more rigorous, task-driven evaluation methodologies that can provide clearer insights into the impact of different design choices in point cloud visualization.

{\bf Task-efficiency vs Computational efficiency.} Additionally, our analysis underscores the importance of evaluating the efficiency of human users in performing point cloud-related tasks. Many techniques introduce significant computational overhead, yet the actual benefits for human users are often unquantified. Understanding the trade-offs between computational complexity and user performance is crucial for the design phase, as some methods may impose unnecessary processing costs without delivering tangible improvements in task completion. Future research should focus on quantifying these trade-offs and ensuring that visualization techniques balance both human needs and computational efficiency.

{\bf Machine vs. Human Consumption.} A final observation is about the lack of clarity regarding whether a point cloud technique is developed for human interpretation or machine processing. While these methods produce visual representations, it is often assumed that they are designed for human users. However, a deeper review of the literature reveals that many techniques seem primarily optimized for machine processing and computational efficiency, with little consideration given to how well they support human interpretation. Highly optimized techniques may not always be necessary though. As an example, we have seen how empirical studies have shown that humans can recognize objects in highly noisy point clouds \cite{gruchalla_structure_2021}, well beyond the tolerance of most machine learning models. This suggests that many filtering techniques are primarily designed to improve machine performance rather than facilitate human perception. If visualization techniques are intended to support human analysis, more research is needed to ensure they effectively balance computational demands with human perceptual requirements.

\section{Limitation of our Study}
\label{sec:limitation}

We acknowledge two limitations in our taxonomy work.

First, First, we acknowledge our review may not be exhaustive regarding applications. Point cloud data is used across numerous disciplines, with a vast body of literature spanning diverse research communities. Given this, it is possible that certain niche applications or emerging trends may not be fully captured. Nevertheless, our approach remains robust. We systematically analyzed a wide range of application areas, identifying and reviewing over 40 papers that explicitly discuss user objectives until we were not able to identify new low-level tasks. We feel this process allowed us to derive the task characterization presented in Section \ref{sec:task}, in a principled manner.

Second, we have deliberately chosen not to include point cloud interaction techniques in our taxonomy, despite their significance. Tasks such as annotation \cite{Franzluebbers2022a}, segmentation \cite{LingyunYu2012}, and direct manipulation of point clouds require dedicated visualization strategies, but we believe these aspects warrant a separate taxonomy. Interaction-focused studies often revolve around input modalities, user interfaces, and display technologies (e.g., VR/AR environments \cite{Elmqvist2006} or immersive multidisplays \cite{Hanula2015}), which introduce distinct considerations beyond visualization alone. Similarly, we have not emphasized display types in our taxonomy, as many works involving alternative display technologies inherently incorporate interactive techniques rather than purely visualization-driven tasks.

\section{Conclusion}

In conclusion, this paper introduces a task-oriented taxonomy for spatial point cloud data visualization, providing a structured framework for categorizing existing techniques based on key dimensions such as data types, intended tasks, and design choices. By systematically analyzing the strengths and limitations of current approaches, we have identified critical challenges, including the need for perception-aware design strategies and low-level task-oriented evaluations. Addressing these challenges is essential to advancing the field and enabling more effective and intuitive visualization tools for users working with complex spatial data.

We believe this taxonomy will not only serve as a foundation for understanding the current landscape of point cloud visualization but also guide future innovations. By refining visualization techniques to be more task-specific and perception-driven, we can better empower users to derive meaningful insights from spatial data.

% \input{sections/Denoising}
% \input{sections/derived-information}
% \input{sections/Rendering}
% \input{sections/overview}

% \input{sections/vis-techniques}
%\input{sections/conclusions}

%% if specified like this the section will be omitted in review mode
%\acknowledgments{%
	%The authors wish to thank A, B, and C.
  %This work was supported in part by a %grant from XYZ (\# 12345-67890).%
%}

\bibliographystyle{abbrv-doi-hyperref}
\bibliography{references,tasks_biblio,taxonomies}
\end{document}